\documentclass[12pt]{article}\usepackage[]{graphicx}\usepackage[]{color}
\makeatletter
\def\maxwidth{ %
  \ifdim\Gin@nat@width>\linewidth
    \linewidth
  \else
    \Gin@nat@width
  \fi
}
\makeatother

\definecolor{fgcolor}{rgb}{0.345, 0.345, 0.345}
\newcommand{\hlnum}[1]{\textcolor[rgb]{0.686,0.059,0.569}{#1}}%
\newcommand{\hlstr}[1]{\textcolor[rgb]{0.192,0.494,0.8}{#1}}%
\newcommand{\hlcom}[1]{\textcolor[rgb]{0.678,0.584,0.686}{\textit{#1}}}%
\newcommand{\hlopt}[1]{\textcolor[rgb]{0,0,0}{#1}}%
\newcommand{\hlstd}[1]{\textcolor[rgb]{0.345,0.345,0.345}{#1}}%
\newcommand{\hlkwa}[1]{\textcolor[rgb]{0.161,0.373,0.58}{\textbf{#1}}}%
\newcommand{\hlkwb}[1]{\textcolor[rgb]{0.69,0.353,0.396}{#1}}%
\newcommand{\hlkwc}[1]{\textcolor[rgb]{0.333,0.667,0.333}{#1}}%
\newcommand{\hlkwd}[1]{\textcolor[rgb]{0.737,0.353,0.396}{\textbf{#1}}}%

\usepackage{framed}
\makeatletter
\newenvironment{kframe}{%
 \def\at@end@of@kframe{}%
 \ifinner\ifhmode%
  \def\at@end@of@kframe{\end{minipage}}%
  \begin{minipage}{\columnwidth}%
 \fi\fi%
 \def\FrameCommand##1{\hskip\@totalleftmargin \hskip-\fboxsep
 \colorbox{shadecolor}{##1}\hskip-\fboxsep
     \hskip-\linewidth \hskip-\@totalleftmargin \hskip\columnwidth}%
 \MakeFramed {\advance\hsize-\width
   \@totalleftmargin\z@ \linewidth\hsize
   \@setminipage}}%
 {\par\unskip\endMakeFramed%
 \at@end@of@kframe}
\makeatother

\definecolor{shadecolor}{rgb}{.97, .97, .97}
\definecolor{messagecolor}{rgb}{0, 0, 0}
\definecolor{warningcolor}{rgb}{1, 0, 1}
\definecolor{errorcolor}{rgb}{1, 0, 0}
\newenvironment{knitrout}{}{} 

\usepackage{alltt}
\usepackage{amsthm,amsmath,amssymb}
\usepackage{alltt}%
\usepackage{hyperref}
\usepackage{makeidx,epsfig}
\usepackage{graphicx,psfrag,epsf}
\usepackage{enumerate}
\usepackage{natbib}
\usepackage{setspace}
\usepackage{xspace,xcolor,soul}

\newcommand{\blind}{0}

\newcommand{\pkg}[1]{\hlstd{\texttt{#1}}}
\newcommand{\argument}[1]{\hlkwc{\texttt{#1}}}
\newcommand{\func}[1]{\hlkwd{\texttt{#1}()}}

\newcommand{\val}[1]{\hlnum{\texttt{#1}}}
\newcommand{\charvec}[1]{\hlstr{\texttt{#1}}}
\newcommand{\data}[1]{\hlstd{\texttt{#1}}}
\newcommand{\cmd}[1]{\hlkwa{\texttt{#1}}}
\newcommand{\obj}[1]{\hlstd{\texttt{#1}}}
\newcommand{\R}{\textsf{R}\xspace}

\author{Benjamin S. Baumer~\thanks{, , Northampton, MA 01063. }}
\IfFileExists{upquote.sty}{\usepackage{upquote}}{}
\begin{document}

\def\spacingset#1{\renewcommand{\baselinestretch}%
{#1}\small\normalsize} \spacingset{1}


\if0\blind
{
  \title{\bf A Grammar for Reproducible and Painless Extract-Transform-Load Operations\\on Medium Data}
  \author{Benjamin S. Baumer\thanks{
    The author gratefully acknowledges the editor, associate editor, and two anonymous reviewers for helpful comments, as well as Carson Sievert, Nicholas Horton, Weijia Zhang, Wencong Li, Rose Gueth, Trang Le, and Eva Gjekmarkaj for their contributions to this work. Email: \texttt{bbaumer@smith.edu}}\hspace{.2cm}\\
    Program in Statistical and Data Sciences, Smith College}
  \maketitle
} \fi

\if1\blind
{
  \bigskip
  \bigskip
  \bigskip
  \begin{center}
    {\LARGE\bf A Grammar for Reproducible and Painless Extract-Transform-Load Operations\\on Medium Data}
\end{center}
  \medskip
} \fi

\bigskip
\begin{abstract}
Many interesting data sets available on the Internet are of a \emph{medium} size---too big to fit into a personal computer's memory, but not so large that they won't fit comfortably on its hard disk. In the coming years, data sets of this magnitude will inform vital research in a wide array of application domains. However, due to a variety of constraints they are cumbersome to ingest, wrangle, analyze, and share in a reproducible fashion. These obstructions hamper thorough peer-review and thus disrupt the forward progress of science. We propose a predictable and pipeable framework for \R (the state-of-the-art statistical computing environment) that leverages SQL (the venerable database architecture and query language) to make reproducible research on medium data a painless reality.  
\end{abstract}

\noindent%
\vfill

\newpage
\spacingset{1.45} 

\section{Introduction}

\subsection{Motivation}

Scientific research is increasingly driven by ``large" data sets. However, the definition of ``large" is relative. We define \emph{medium} data to be those who are too big to store in memory on a personal computer, but not so big that they won't fit on a hard drive. Typically, this means data on the order of several gigabytes (see Table~\ref{tab:data}).
Publicly accessible medium data sets (PAMDAS) are now available in a variety of application domains. A few examples are the Citi Bike bike-sharing program in New York City, campaign contributions from the Federal Election Commission, and on-time airline records from the Bureau of Transportation Statistics. 

\begin{table}[b]
  \centering
  \begin{tabular}{cccc}
``Size" & actual size & hardware & software \\
\hline
small  & $<$ several GB & RAM & \R \\ 
medium & several GB -- a few TB & hard disk & SQL \\
big    & many TB or more & computing cluster & Spark? \\
\hline
  \end{tabular}
  \caption{Relative sizes of data from the point-of-view of personal computer users. We focus on medium data, which are too large to fit comfortably into the memory of a typical personal computer, but not so large that they won't fit comfortably on the hard drive of such a computer. In 2018, desktop computers typically ship with hard drives of at most four terabytes. Most laptops use solid-state hard drives which hold less than one terabyte. \label{tab:data}}
\end{table}

While PAMDAS provide \emph{access}, they do not remove all barriers to reproducible research on these data. Because of their size, reading these data into memory will either take an unbearably long time, exhaust the computer's memory until it grinds to a halt, or simply not work at all. A sensible solution is to download the data to a local storage device and then import it into a relational database management system (RDBMS). RDBMS's have been around since the 1970s, and provide a scalable solution for data of this magnitude. High-quality, open source implementations (e.g., MySQL, PostgreSQL, SQLite) are prevalent. However, creating a new database from scratch is time-consuming and requires knowledge of database administration. While these skills are not difficult to acquire, they are not always emphasized in the traditional undergraduate curriculum in either statistics~\citep{asa-guidelines} or computer science~\citep{acm2013guidlines}. 

The process of downloading the raw data from its authoritative source and importing it into a RDBMS is often called \emph{Extract-Transform-Load} (ETL). Professionals who work with data spend a disproportionate amount of their time on such tasks. Their solutions are sometimes idiosyncratic, platform- or architecture-specific, poorly documented, unshared, and involve custom scripts written in various (and often multiple) languages. Thus, the ETL process constitutes a barrier to reproducible research, because there is not a good way of verifying that two people downloading data from the same source will end up with the \emph{exact} same set of data in their local data store. This matters because any subsequent data analysis could be sensitive to small perturbations in the underlying data set. Like~\cite{claerbout1994hypertext} and~\cite{donoho2010invitation}, we recognize the necessity of data-based research being backed by open data, with well-documented data analysis code that is shared publicly and executable on open-source platforms.

Sharing the local data store is often also problematic, due to licensing restrictions and the sheer size of the data. While PAMDAS may be free to download, there may be legal barriers to publicly sharing a local data store that is essentially a reproduction of those original data (see, for example~\cite{greenhouse2008}). File size limitations imposed by software repositories (e.g., GitHub, CRAN) may make distribution via those channels unfeasible. Moreover, sharing medium data sets through the cloud may be expensive or unrealistic for many individuals and small companies. 

\subsection{Our contribution}

We propose a software framework for simultaneously solving two related but distinct problems when analyzing PAMDAS: 1) how to build a relational database with little or no knowledge of SQL, and; 2) how to ensure reproducibility in published research succinctly. Our solution consists of a package for \R~\citep{rcran} that provides a \emph{core} framework for ETL operations along with a series of \emph{peripheral} packages that extend the ETL framework for a specific PAMDAS. The core \pkg{etl} package is available on CRAN~\citep{etl}. Seven different peripheral packages are in various states of development, but in principle there is no limit to how many could be created. These packages are all fully open source (hosted on GitHub with Creative Commons licenses), cross-platform (to the extent allowed by \R, MySQL, PostgreSQL, SQLite, etc.), and fit into the state-of-the-art \pkg{tidyverse}~\citep{tidyverse} paradigm popular among \R users. Specifically, the \pkg{etl} package extends the functionality of existing database tools in \R (see Section~\ref{sec:database}) by maintaining local storage locations and employing a consistent grammar for ETL operations (see Section~\ref{sec:grammar}). 

The \pkg{etl} suite of packages will make it easier to bring PAMDAS to data analysts of all stripes while lowering barriers to entry and enhancing transparency, usability, and reproducibility. For the most part, knowledge of SQL will not be required to access these data through \R. (See \cite{kline2005sql} for a primer on SQL.)

In Section~\ref{sec:examples}, we provide motivating examples that illustrate how the use of the \pkg{etl} framework can facilitate the construction of medium databases for PAMDAS, and how that ability can improve reproducibility in published research. We explicate the grammar employed by \pkg{etl}---and how it speeds adoption---in Section~\ref{sec:grammar}. In Section~\ref{sec:user}, we describe how a typical \R user can use the \pkg{etl} package and its dependent packages to build medium databases with relative ease. In Section~\ref{sec:devel}, we briefly outline how an \R developer can rapidly create their own \pkg{etl}-dependent packages. We conclude with a brief discussion in Section~\ref{sec:conclusion}. In our supplementary materials, Section~\ref{sec:related} situates our work in the existing ecosystem of \R tools, Section~\ref{sec:mtcars} provides a short example of how to use \pkg{etl}, Section~\ref{sec:benchmark} discusses performance benchmarks, and Section~\ref{sec:amazon} illustrates how cloud computing services can be used in conjunction with the \pkg{etl} package. 

\section{Motivating examples}
\label{sec:examples}

\subsection{ETL workflow for on-time airline data}
\label{sec:airlines}

The \pkg{etl} package provides the foundation for \pkg{etl}-dependent packages that focus on specific data sets. In this example, we illustrate how one of these packages---\pkg{airlines}---can be used to build a medium database of flight information. These data are available from the Bureau of Transportation Statistics via monthly ZIP files. 

First, we load the \pkg{airlines} package. 
We then use the \func{src\_mysql\_cnf} function provided by \pkg{etl} to create a database connection to a preconfigured remote MySQL~\footnote{MySQL is a popular open-source relational database management system. The \func{src\_mysql\_cnf} function reads server information and credentials from a configuration file stored in the user's home directory.} database server~\footnote{The command \cmd{library(airlines)} makes user-facing functions from both the \pkg{airlines} and \pkg{etl} packages available.}. (The database to which we connect is also called \charvec{"airlines"}.)


\begin{knitrout}
\definecolor{shadecolor}{rgb}{0.969, 0.969, 0.969}\color{fgcolor}\begin{kframe}
\begin{alltt}
\hlkwd{library}\hlstd{(airlines)}
\hlstd{db} \hlkwb{<-} \hlkwd{src_mysql_cnf}\hlstd{(}\hlstr{"airlines"}\hlstd{,} \hlkwc{groups} \hlstd{=} \hlstr{"scidb"}\hlstd{)}
\end{alltt}
\end{kframe}
\end{knitrout}

Next, we instantiate an object called \obj{ontime}. The source of \obj{ontime}'s data is the \R package called \charvec{"airlines"}.
In this case, we specify the \argument{db} argument to be the connection to our MySQL database, and the \argument{dir} argument for local storage. Any files we download or transform will be stored in \argument{dir}. 
Among other things, \obj{ontime} is a \cmd{src\_dbi} object---an interface to a database---meaning that it can take advantage of the many functions---mainly provided by the \pkg{dbplyr}~\citep{dbplyr} package---that work on such objects. We postpone a more detailed discussion of this until Section~\ref{sec:nouns}.

\begin{knitrout}
\definecolor{shadecolor}{rgb}{0.969, 0.969, 0.969}\color{fgcolor}\begin{kframe}
\begin{alltt}
\hlstd{ontime} \hlkwb{<-} \hlkwd{etl}\hlstd{(}\hlstr{"airlines"}\hlstd{,} \hlkwc{db} \hlstd{= db,} \hlkwc{dir} \hlstd{=} \hlstr{"~/dumps/airlines"}\hlstd{)}
\end{alltt}
\end{kframe}
\end{knitrout}

We then perform our ETL operations. We first initialize the database with \func{etl\_init}, which in this case loads table schemas from an SQL script provided by the \pkg{airlines} package. Next, the \func{etl\_extract} function downloads data from 1987--2016. This results in one ZIP file for each month being stored in a subdirectory of \argument{dir}. Next, we use the \func{etl\_transform} function to unzip these files and grab the relevant CSVs~\footnote{CSV stands for \href{https://en.wikipedia.org/wiki/Comma-separated_values}{Comma-Separated Values}, and is a common data format.}. While the \func{etl\_transform} function takes the same arguments as \func{etl\_extract}, those arguments needn't take the same values. For purposes of illustration we choose to transform only the data from the decade of the 1990s. Finally, the \func{etl\_load} function reads the CSV data from 1996 and 1997 into the database, but only from the first half of the year, plus September. 

\begin{knitrout}
\definecolor{shadecolor}{rgb}{0.969, 0.969, 0.969}\color{fgcolor}\begin{kframe}
\begin{alltt}
\hlstd{ontime} \hlopt{%>%}
  \hlkwd{etl_init}\hlstd{()} \hlopt{%>%}
  \hlkwd{etl_extract}\hlstd{(}\hlkwc{years} \hlstd{=} \hlnum{1987}\hlopt{:}\hlnum{2016}\hlstd{)} \hlopt{%>%}
  \hlkwd{etl_transform}\hlstd{(}\hlkwc{years} \hlstd{=} \hlnum{1990}\hlopt{:}\hlnum{1999}\hlstd{)} \hlopt{%>%}
  \hlkwd{etl_load}\hlstd{(}\hlkwc{years} \hlstd{=} \hlnum{1996}\hlopt{:}\hlnum{1997}\hlstd{,} \hlkwc{months} \hlstd{=} \hlkwd{c}\hlstd{(}\hlnum{1}\hlopt{:}\hlnum{6}\hlstd{,} \hlnum{9}\hlstd{))}
\end{alltt}
\end{kframe}
\end{knitrout}

We note that this process---which may take several hours---results in a relational database with multiple tables occupying several dozen gigabytes on disk, and comprising several million rows of data (the full data set across all years contains more than 160 million rows). 

\begin{knitrout}
\definecolor{shadecolor}{rgb}{0.969, 0.969, 0.969}\color{fgcolor}\begin{kframe}
\begin{alltt}
\hlstd{ontime}
\end{alltt}
\begin{verbatim}
## dir:  728 files occupying 26.255 GB
## src:  mysql 5.5.58-0ubuntu0.14.04.1-log [bbaumer@scidb.smith.edu:/airlines]
## tbls: airports, carriers, flights, planes, summary, weather
\end{verbatim}
\end{kframe}
\end{knitrout}

Moreover, \obj{ontime} is constructed such that we can use existing functionality from the \pkg{dplyr} package~\citep{dplyr} to access the data from a specific airport, say, Bradley International (BDL), which serves Hartford, CT and Springfield, MA. Please see Section~\ref{sec:nouns} for more technical details. 

\begin{knitrout}
\definecolor{shadecolor}{rgb}{0.969, 0.969, 0.969}\color{fgcolor}\begin{kframe}
\begin{alltt}
\hlstd{ontime} \hlopt{%>%}
  \hlkwd{tbl}\hlstd{(}\hlstr{"flights"}\hlstd{)} \hlopt{%>%}
  \hlkwd{filter}\hlstd{(year} \hlopt{==} \hlnum{1996}\hlstd{, dest} \hlopt{==} \hlstr{"BDL"}\hlstd{)} \hlopt{%>%}
  \hlkwd{head}\hlstd{(}\hlnum{3}\hlstd{)}
\end{alltt}
\begin{verbatim}
## # Source:   lazy query [?? x 21]
## # Database: mysql 5.5.58-0ubuntu0.14.04.1-log
## #   [bbaumer@scidb.smith.edu:/airlines]
##    year month   day dep_time sched_dep_time dep_delay arr_time
##   <int> <int> <int>    <int>          <int>     <int>    <int>
## 1  1996    10     1      637            640        -3      940
## 2  1996    10     1      652            653        -1     1106
## 3  1996    10     1      707            710        -3      836
## # ... with 14 more variables: sched_arr_time <int>, arr_delay <int>,
## #   carrier <chr>, tailnum <chr>, flight <int>, origin <chr>, dest <chr>,
## #   air_time <int>, distance <int>, cancelled <int>, diverted <int>,
## #   hour <int>, minute <int>, time_hour <chr>
\end{verbatim}
\end{kframe}
\end{knitrout}

Thus, with just a few simple lines of \R code---and no knowledge of SQL---the \pkg{airlines} package allows us to create a medium-sized relational database suitable for analysis with popular \pkg{dplyr} tools. 

\subsection{Reproducible research with Citi Bike data}
\label{sec:bikes}

The following example using the \pkg{citibike} package illustrates how reproducibility of published research in the natural and social sciences could be improved through use of the \pkg{etl} framework. 

The lack of reproducibility in published scientific research in the natural and social sciences is problematic. Here, we revisit a series of operations research efforts analyzing load balancing for stations in the Citi Bike municipal bike sharing system in New York City~\citep{o2015data,singhvi2015predicting,o2015smarter} and demonstrate how the \pkg{etl} framework improves data analytic workflows. The data from this system has fueled several research efforts since its launch in July 2013. 

The system's engineers face a problem balancing the load of bikes among stations. Since one cannot ensure that bikes rented from one station will be returned to that station, how can one ensure that there will always be enough bikes at a particular station to meet demand? 

\cite{singhvi2015predicting} provided the following description of their data set:
\begin{quotation}
We obtained bike usage statistics for April, May, June and July 2014 from Citi Bike's website (\url{https://www.citibikenyc.com/system-data}). This dataset contains start station id, end station id, station latitude, station longitude and trip time for each bike trip. 332 bike stations have one or more originating bike trips. 253 of these are in Manhattan while 79 are in Brooklyn (left panel of Figure 1). We processed this raw data to get the number of bike trips between each station pair during morning rush hours.
\end{quotation}

This is a fairly specific description of how the data were acquired, since it cites a URL, a specific date range, and the exact number of stations present. However, is it sufficient information for someone else to verify that they are working with the same data set?

Using the \pkg{citibike} package, we attempt to reproduce this data set by creating a connection to a (in this case local) database, initializing it, and then populating that database with a single call to \func{etl\_update}:

\begin{knitrout}
\definecolor{shadecolor}{rgb}{0.969, 0.969, 0.969}\color{fgcolor}\begin{kframe}
\begin{alltt}
\hlkwd{library}\hlstd{(citibike)}
\hlstd{bikes} \hlkwb{<-} \hlkwd{etl}\hlstd{(}\hlstr{"citibike"}\hlstd{,} \hlkwc{dir} \hlstd{=} \hlstr{"~/dumps/citibike/"}\hlstd{,}
             \hlkwc{db} \hlstd{=} \hlkwd{src_mysql_cnf}\hlstd{(}\hlstr{"citibike"}\hlstd{))}
\end{alltt}
\end{kframe}
\end{knitrout}

\begin{knitrout}
\definecolor{shadecolor}{rgb}{0.969, 0.969, 0.969}\color{fgcolor}\begin{kframe}
\begin{alltt}
\hlstd{bikes} \hlopt{%>%}
  \hlkwd{etl_update}\hlstd{(}\hlkwc{years} \hlstd{=} \hlnum{2014}\hlstd{,} \hlkwc{months} \hlstd{=} \hlnum{4}\hlopt{:}\hlnum{7}\hlstd{)}
\end{alltt}
\end{kframe}
\end{knitrout}

Leveraging \pkg{dplyr} again, the following pipeline confirms the number of unique stations. 

\begin{knitrout}
\definecolor{shadecolor}{rgb}{0.969, 0.969, 0.969}\color{fgcolor}\begin{kframe}
\begin{alltt}
\hlstd{trips} \hlkwb{<-} \hlstd{bikes} \hlopt{%>%}
  \hlkwd{tbl}\hlstd{(}\hlstr{"trips"}\hlstd{)}
\hlstd{trips} \hlopt{%>%}
  \hlkwd{group_by}\hlstd{(Start_Station_ID)} \hlopt{%>%}
  \hlkwd{summarize}\hlstd{(}\hlkwc{num_trips} \hlstd{=} \hlkwd{n}\hlstd{())} \hlopt{%>%}
  \hlkwd{filter}\hlstd{(num_trips} \hlopt{>=} \hlnum{1}\hlstd{)} \hlopt{%>%}
  \hlkwd{collect}\hlstd{()} \hlopt{%>%}
  \hlkwd{nrow}\hlstd{()}
\end{alltt}
\begin{verbatim}
## [1] 332
\end{verbatim}
\end{kframe}
\end{knitrout}

How confident are you that we now have a copy of the same data as these researchers? We have the same number of stations, but do we have the same number of rows? Do the rows contain the same information? These questions are impossible to verify given the description above.

Behind the scenes, the authors certainly wrote code to download and process these data from the Citi Bike website. Indeed, they admit as much in the last sentence of the quotation above. Moreover, the figures in the paper were clearly produced in \R. Thus, this research provides a perfect instance where the use of the \pkg{citibike} package could have standardized the exact data set upon which their research is based. The inclusion of a few short lines of code would ensure that all parties are analyzing the same data set. 

In another effort, \cite{faghih2016incorporating} model bike demand using spatio-temporal data from the Citi Bike system. Their description of the data is less specific than that of~\cite{singhvi2015predicting}, however they include an appendix containing some summary statistics. There is no clear way to verify the integrity of the data set. They write:
\begin{quotation}
We focused on the month of September, 2013; i.e. the peak month of the usage in 2013. Therefore, the final sample consists of 237,600 records (330 stations $\times$ 24 hours $\times$ 30 days).
\end{quotation}

Here again, a single call to \func{etl\_update} could have ensured that all users have the same data set:

\begin{knitrout}
\definecolor{shadecolor}{rgb}{0.969, 0.969, 0.969}\color{fgcolor}\begin{kframe}
\begin{alltt}
\hlkwd{etl_update}\hlstd{(bikes,} \hlkwc{year} \hlstd{=} \hlnum{2013}\hlstd{,} \hlkwc{months} \hlstd{=} \hlnum{9}\hlstd{)}
\end{alltt}
\end{kframe}
\end{knitrout}

The number of records reported is somewhat misleading, since many stations had no trips during some hours of some day. In fact, the following pipeline returns only $167,258$ records. 

\begin{knitrout}
\definecolor{shadecolor}{rgb}{0.969, 0.969, 0.969}\color{fgcolor}\begin{kframe}
\begin{alltt}
\hlstd{trips} \hlopt{%>%}
  \hlkwd{filter}\hlstd{(}\hlkwd{YEAR}\hlstd{(Start_Time)} \hlopt{==} \hlnum{2013}\hlstd{)} \hlopt{%>%}
  \hlkwd{group_by}\hlstd{(Start_Station_ID,} \hlkwd{DAY}\hlstd{(Start_Time),} \hlkwd{HOUR}\hlstd{(Start_Time))} \hlopt{%>%}
  \hlkwd{summarize}\hlstd{(}\hlkwc{N} \hlstd{=} \hlkwd{n}\hlstd{(),}
            \hlkwc{num_stations} \hlstd{=} \hlkwd{COUNT}\hlstd{(}\hlkwd{DISTINCT}\hlstd{(Start_Station_ID)),}
            \hlkwc{num_days} \hlstd{=} \hlkwd{COUNT}\hlstd{(}\hlkwd{DISTINCT}\hlstd{(}\hlkwd{DAYOFYEAR}\hlstd{(Start_Time))))} \hlopt{%>%}
  \hlkwd{collect}\hlstd{()} \hlopt{%>%}
  \hlkwd{nrow}\hlstd{()}
\end{alltt}
\begin{verbatim}
## [1] 167258
\end{verbatim}
\end{kframe}
\end{knitrout}


In both cases, our attempt to verify the data used by these researchers was greatly aided by the \pkg{citibike} package. Moreover, because the \pkg{citibike} package employs a consistent grammar and fits into the popular \pkg{tidyverse}, it is far easier to use than say, a \cmd{bash} script posted on one of these researchers' website.

\section{A grammar for ETL}
\label{sec:grammar}

While the individual steps necessary to process a PAMDAS into a RDBMS vary greatly, the three major steps of downloading the data, wrangling it, and importing it into a database are universal. The \pkg{etl} framework is designed to take advantage of this common structure. This achieves two major goals: to abstract the idiosyncratic complications of each PAMDAS away from the user, and; to restrict the developer's obligation to only those idiosyncracies. 

The use of the term ``grammar" in a data science context is not novel. \cite{wilkinson2006grammar} described a ``grammar of graphics" that was implemented in \R as \pkg{ggplot2}~\citep{ggplot2}. Similarly, \pkg{dplyr}~\citep{dplyr} provides a ``grammar of data manipulation." A grammar consists of verbs and nouns that can be combined in logical ways to intentional effect. The benefit of having one is that once a user understands the grammar, they should be able to read and write longer sequences of code fluently. The use of the pipe operator provided by the \pkg{magrittr} package~\citep{magrittr} is crucial to allowing pipelines (i.e., ``sentences") to be composed from short sequences of commands (i.e., ``phrases").

The design of \pkg{etl} is very much in this spirit---it is an extension of the grammar of data manipulation provided by \pkg{dplyr}. We present \pkg{etl} as a grammar for ETL operations that is rich enough to describe a great many ETL processes, but simple enough to contain only a handful of verbs. In Section~\ref{sec:use_cases}, we illustrate how different ETL ``sentences" can cover several common use cases. These cases are informed by our experience working with data of this magnitude in a variety of professional contexts over the past 15 years.

\subsection{Tidyverse design}

The \pkg{etl} package fits into a growing collection of \R packages known as the \pkg{tidyverse}~\citep{tidyverse}. These packages are designed for interoperability and emphasize functions that are \href{https://www.meetup.com/nyhackr/events/224749681/}{\emph{pure}}, \emph{predictable}, and \href{http://newyorktechjournal.com/2015/09/pure-predictable-pipeable-creating-fluent-interfaces-with-r/}{\emph{pipeable}}, as described by Hadley Wickham.

\begin{description}
  \item[Pure] The output of a function is entirely dependent on the input to the function. Pure functions make no changes to other objects in the environment.
  \item[Predictable] Functions names, arguments, and behaviors are consistent, such that if you can learn how to use one function, you have a head start on understanding how to use others. 
  \item[Pipeable] Functions return objects of the same type as their first argument, so that pipeable operations can be chained together to produce \emph{pipelines}.
\end{description}

Functions in the \pkg{etl} package are predictable and pipeable, but not pure. This is by design---while the predictability and pipeability make \pkg{etl} easy to use and compatible with the \pkg{tidyverse}, these functions also necessarily download files, store them locally, and interact with databases outside of \R. These changes to the computing environment are unavoidable given the nature of the task.

\subsection{ETL nouns}
\label{sec:nouns}

At the center of any \pkg{etl} pipeline is an object that is created by the \func{etl} function, whose first argument is a character string naming the package that provides access to the data. The package \pkg{foo} creates objects of class \cmd{etl\_foo}. If it is not installed, \func{etl} will throw an error. 

\begin{knitrout}
\definecolor{shadecolor}{rgb}{0.969, 0.969, 0.969}\color{fgcolor}\begin{kframe}
\begin{alltt}
\hlkwd{etl}\hlstd{(}\hlstr{"nyctaxi"}\hlstd{)}
\end{alltt}

{\ttfamily\noindent\itshape\color{messagecolor}{\#\# No database was specified so I created one for you at:}}

{\ttfamily\noindent\itshape\color{messagecolor}{\#\# /tmp/Rtmpdcqje1/file4fd51f7d8e53.sqlite3}}\begin{verbatim}
## dir:  0 files occupying 0 GB
## src:  sqlite 3.22.0 [/tmp/Rtmpdcqje1/file4fd51f7d8e53.sqlite3]
## tbls:
\end{verbatim}
\begin{alltt}
\hlkwd{etl}\hlstd{(}\hlstr{"foo"}\hlstd{)}
\end{alltt}

{\ttfamily\noindent\bfseries\color{errorcolor}{\#\# Error in etl.default("{}foo"{}): Please make sure that the 'foo' package is installed}}\end{kframe}
\end{knitrout}

Recall that all \pkg{etl} objects are \cmd{src\_dbi} objects. Thus, \func{print}, \func{summary}, and \func{is} methods for \cmd{etl} objects extend those provided by other packages. Here, we illustrate a few of these features. 

\begin{knitrout}
\definecolor{shadecolor}{rgb}{0.969, 0.969, 0.969}\color{fgcolor}\begin{kframe}
\begin{alltt}
\hlkwd{class}\hlstd{(ontime)}
\end{alltt}
\begin{verbatim}
## [1] "etl_airlines" "etl"          "src_dbi"      "src_sql"     
## [5] "src"
\end{verbatim}
\begin{alltt}
\hlcom{# summary(ontime)  }
\hlcom{# output suppressed for space}
\hlkwd{src_tbls}\hlstd{(ontime)}
\end{alltt}
\begin{verbatim}
## [1] "airports" "carriers" "flights"  "planes"   "summary"  "weather"
\end{verbatim}
\end{kframe}
\end{knitrout}

Moreover, like all \cmd{src\_dbi} objects, every \pkg{etl} object is stored as a list and maintains a \cmd{DBIConnection} to a database in \val{con}.

\begin{knitrout}
\definecolor{shadecolor}{rgb}{0.969, 0.969, 0.969}\color{fgcolor}\begin{kframe}
\begin{alltt}
\hlkwd{str}\hlstd{(ontime)}
\end{alltt}
\begin{verbatim}
## List of 2
##  $ con  :Formal class 'MySQLConnection' [package "RMySQL"] with 1 slot
##   .. ..@ Id: int [1:2] 0 0
##  $ disco:<environment: 0x53b1638> 
##  - attr(*, "class")= chr [1:5] "etl_airlines" "etl" "src_dbi" "src_sql" ...
##  - attr(*, "pkg")= chr "airlines"
##  - attr(*, "dir")= chr "/home/bbaumer/dumps/airlines"
##  - attr(*, "raw_dir")= chr "/home/bbaumer/dumps/airlines/raw"
##  - attr(*, "load_dir")= chr "/home/bbaumer/dumps/airlines/load"
\end{verbatim}
\end{kframe}
\end{knitrout}

Accessing this \val{con} allows one to make use of the extensive functionality provided by the \pkg{DBI}~\citep{DBI} package. 

\begin{knitrout}
\definecolor{shadecolor}{rgb}{0.969, 0.969, 0.969}\color{fgcolor}\begin{kframe}
\begin{alltt}
\hlstd{DBI}\hlopt{::}\hlkwd{dbGetInfo}\hlstd{(ontime}\hlopt{$}\hlstd{con)}
\hlstd{DBI}\hlopt{::}\hlkwd{dbListTables}\hlstd{(ontime}\hlopt{$}\hlstd{con)}
\end{alltt}
\end{kframe}
\end{knitrout}

\pkg{etl} objects can interface with any RDBMS that can become a \cmd{src\_dbi}. In particular, in addition to SQLite, MySQL/MariaDB, PostgreSQL, Google BigQuery, and MonetDB---all of which are supported through standalone \R packages---\pkg{DBI} functionality can be used with a variety of other RDBMSs through the \pkg{odbc} package~\citep{odbc}, which supports Amazon Redshift, Apache Hive, Apache Impala, Microsoft SQL Server, Oracle, Salesforce, and Teradata.~\footnote{See \url{https://db.rstudio.com/databases/} for the most current list.} For those that lack a supportive IT infrastructure, low cost, low maintenance access to many of these technologies is available through cloud computing vendors, such as \href{https://aws.amazon.com/rds/}{Amazon RDS} (see Appendix~\ref{sec:amazon} for a brief tutorial).

The main difference between an \pkg{etl} object and a \cmd{src\_dbi} object is that an \pkg{etl} object has attributes that point toward \cmd{dir}---a directory where files can be safely read and written. If no \cmd{dir} argument is specified, a temporary directory is created and used. Within \cmd{dir}, two subdirectories are automatically created: \cmd{raw} and \cmd{load}. Raw files downloaded via \func{etl\_extract} are placed in \cmd{raw}. \func{etl\_transform} reads those files and writes the resulting transformed files to \cmd{load}. Finally, the \func{etl\_load} function reads files from \cmd{load} and imports them into the database.

\subsection{ETL verbs}

The workhorses of \pkg{etl} are the three main verbs. Each takes an \pkg{etl} object as its first argument and returns an \pkg{etl} object invisibly, enabling these functions to be piped. 

\begin{itemize}
  \item \func{etl\_extract}: download data from the Internet and place the raw files in the \cmd{raw} directory. The \cmd{default} method grabs data provided by the named package. 
  \item \func{etl\_transform}: read files in the \cmd{raw} directory, perform any necessary data wrangling operations, and write CSV files to the \cmd{load} directory. The \cmd{default} method copies all CSVs in the \cmd{raw} directory to the \cmd{load} directory. 
  \item \func{etl\_load}: import CSV files from the \cmd{load} directory into the database. The \cmd{default} method imports all CSVs in the \cmd{load} directory into eponymous tables. 
\end{itemize}

Writing these three functions becomes each \pkg{etl}-dependent package maintainer's responsibility. We discuss this in greater detail in Section~\ref{sec:devel}.

While these three main verbs may be the most universal, two other commonly-used verbs are \func{etl\_init} and \func{etl\_cleanup}. 

\begin{itemize}
  \item \func{etl\_init}: initializes the database by either running a SQL initialization script or by simply deleting all of the tables in the existing database. That script can be bundled by the package maintainer or passed as a file path or character vector. It can also be written in generic SQL or in a flavor of SQL specific to a particular database engine. This enables \R users to make use of features that exist in one database implementation but not another (e.g., partitions in MySQL which are not available in SQLite). This step is optional, since \func{DBI::dbWriteTable} will perform column type interpolation during the \func{etl\_load} phase if the corresponding tables don't already exist.
  \item \func{etl\_cleanup}: delete files from either the \cmd{raw} or \cmd{load} directories using regular expression pattern matching. 
\end{itemize}

For convenience, two additional verbs are provided: 

\begin{itemize}
  \item \func{etl\_update}: chains the extract, transform, and load phases together, passing the same arguments to each.
  \item \func{etl\_create}: runs the full chain including initialization, update, and cleanup. 
\end{itemize}

\begin{knitrout}
\definecolor{shadecolor}{rgb}{0.969, 0.969, 0.969}\color{fgcolor}\begin{kframe}
\begin{alltt}
\hlkwd{getS3method}\hlstd{(}\hlstr{"etl_update"}\hlstd{,} \hlstr{"default"}\hlstd{)}
\end{alltt}
\begin{verbatim}
## function(obj, ...) {
##   obj <- obj %>%
##     etl_extract(...) %>%
##     etl_transform(...) %>%
##     etl_load(...)
##   invisible(obj)
## }
## <environment: namespace:etl>
\end{verbatim}
\begin{alltt}
\hlkwd{getS3method}\hlstd{(}\hlstr{"etl_create"}\hlstd{,} \hlstr{"default"}\hlstd{)}
\end{alltt}
\begin{verbatim}
## function(obj, ...) {
##   obj <- obj %>%
##     etl_init(...) %>%
##     etl_update(...) %>%
##     etl_cleanup(...)
##   invisible(obj)
## }
## <environment: namespace:etl>
\end{verbatim}
\end{kframe}
\end{knitrout}

\subsection{Common use cases}
\label{sec:use_cases}

In Section~\ref{sec:bikes}, we showed how a single call to \func{etl\_update} could be used to populate a static database with data specific to a time interval. While this one-shot usage may be the most common, the \pkg{etl} grammar is flexible enough to accommodate other use cases. 

\begin{description}
  \item[Regular updates for updated data] You receive a daily dump of customer data from a vendor in files that are overwritten. Run a script each day that contains \func{etl\_create} to rebuild your database.
  \item[Regular updates for new data] Your web logs are archived into a new file each month. Run \func{etl\_init} once, then run \func{etl\_update} monthly when new data becomes available. 
  \item[Asynchronous updates] A temporary network disruption may result in corrupt downloads or a broken pipeline. The \pkg{etl} design makes it relatively easy to, say, use \func{etl\_cleanup} to delete a single corrupted month of airline data, then re-run \func{etl\_update} on just that month. Since the data are stored in a database, the order of the rows is generally irrelevant. Some care is necessary to avoid duplicate rows, however. 
  \item[Reconfigure and reload] An update to \pkg{airlines} adds a partitioning scheme to the \data{flights} table. You update your database by running \func{update.packages}, followed by \func{etl\_init} and \func{etl\_load}. You do not need to download or transform the data again. 
  \item[Porting a database] You create a local copy of a database, verify its contents, and then port it to a remote server by defining a new database connection, and then calling \func{etl\_load} on the new \pkg{etl} object. 
\end{description}

\section{The \pkg{etl} package for \R users}
\label{sec:user}

The \pkg{etl} framework is designed to make PAMDAS accessible to \R users who may not have experience with SQL. The following \pkg{etl}-dependent packages---which are in various stages of development---can be used in a manner similar to the \pkg{airlines} and \pkg{citibike} packages illustrated in Section~\ref{sec:examples}, since they all employ the grammar described in Section~\ref{sec:grammar}. For further examples, please see Appendix~\ref{sec:mtcars} and the ``\href{https://cran.r-project.org/web/packages/etl/vignettes/using_etl.html}{Using \pkg{etl}}" vignette~\footnote{\url{https://cran.r-project.org/web/packages/etl/vignettes/using_etl.html}}.
These packages---combined with the ability to convert data in any \R package to a relational database as described in Section~\ref{sec:default}---lower barriers of entry to medium data for even novice \R users. 

\subsection{PAMDAS accessible via \pkg{etl}}
\label{sec:PAMDAS}

The following \pkg{etl}-dependent packages are available on GitHub (and CRAN where indicated):

\begin{description}
  \item[macleish] (CRAN) weather and spatial data from the Smith College MacLeish Field Station in Whately, MA~\citep{macleish}
  \item[airlines] on-time flight data from the Bureau of Transportation Services for all domestic flights since October 1987~\citep{airlines}
  \item[imdb] a mirror of the Internet Movie Database~\citep{imdb}
  \item[nyc311] calls to New York City's non-emergency municipal services hotline~\citep{nyc311}
  \item[fec] campaign finance contributions and spending from the Federal Election Commission~\citep{fec}
  \item[citibike] trip data for New York City's municipal bike sharing service~\citep{citibike}
  \item[nyctaxi] (CRAN) trip data from the New York City Taxi and Limousine Commission~\citep{nyctaxi} 
\end{description}

In Section~\ref{sec:devel}, we explain how these packages can be developed rapidly using the \pkg{etl} framework. In some cases, these packages can be reduced to a few lines of \R code.

\subsection{ETL for small data bundled in \R packages}
\label{sec:default}

The \pkg{etl} package can also perform default ETL operations on data stored in any \R package. Here, we build a database of five tables included in the \pkg{nasaweather} package~\citep{nasaweather}. 

\begin{knitrout}
\definecolor{shadecolor}{rgb}{0.969, 0.969, 0.969}\color{fgcolor}\begin{kframe}
\begin{alltt}
\hlstd{nasa} \hlkwb{<-} \hlkwd{etl}\hlstd{(}\hlstr{"nasaweather"}\hlstd{)} \hlopt{%>%}
  \hlkwd{etl_update}\hlstd{()}
\end{alltt}

{\ttfamily\noindent\itshape\color{messagecolor}{\#\# No database was specified so I created one for you at:}}

{\ttfamily\noindent\itshape\color{messagecolor}{\#\# /tmp/Rtmpdcqje1/file4fd559e57b2.sqlite3}}

{\ttfamily\noindent\itshape\color{messagecolor}{\#\# Loading 5 file(s) into the database...}}\begin{alltt}
\hlstd{nasa}
\end{alltt}
\begin{verbatim}
## dir:  10 files occupying 0.008 GB
## src:  sqlite 3.22.0 [/tmp/Rtmpdcqje1/file4fd559e57b2.sqlite3]
## tbls: atmos, borders, elev, glaciers, storms
\end{verbatim}
\end{kframe}
\end{knitrout}

This functionality is a convenience, since data bundled in \R packages are usually small, but it nevertheless allows \R users to create relational databases with minimal effort. 
We note here that since no pre-existing database connection was specified, a SQLite database was created in a temporary directory. 

\section{The \pkg{etl} package for \R developers}
\label{sec:devel}

\subsection{Database functionality in \R}
\label{sec:database}

Recent advances in \R computing have made accessing databases through \R a relatively painless process. 

In \R, a \cmd{data.frame} is a two-dimensional array of data that consists of rows and columns. It is logically analogous to a \emph{table} in SQL parlance, but with two crucial differences in implementation: first, a \cmd{data.frame} is stored in memory, whereas a table is usually written to disk; second, a \cmd{data.frame} need not and cannot be indexed, whereas tables are often indexed. The \pkg{tibble} package in \R extends the \cmd{data.frame} to the more flexible \cmd{tbl} data structure~\citep{tibble}. The \pkg{dbplyr} package further extends the functionality of \cmd{tbl}'s to be backed by a local or remote database~\citep{dbplyr}. A common interface to such databases is provided by the \pkg{DBI} package~\citep{DBI}. Each RDBMS has its own \R package that implements the \pkg{DBI} programming interface. For example, the \pkg{RMySQL} package implements the \pkg{DBI} specification for MySQL~\citep{RMySQL}, while the \pkg{RSQLite} package implements the \pkg{DBI} specification for SQLite~\citep{RSQLite}. 
Through this chain of interfaces, a \cmd{tbl\_mysql} appears to an \R user to be a familiar \cmd{data.frame}, but in fact, it is akin to a \cmd{VIEW} of the underlying MySQL table, and thus occupies virtually no space in \R's memory, and can make use of SQL indexes. 

This infrastructure provides a backdrop for the popular data wrangling package \pkg{dplyr}~\citep{dplyr}, which re-imagines SQL \cmd{SELECT} syntax as a pipeable sequence of data verbs. This approach is attractive because \R users can perform SQL-style operations from within \R without having to learn SQL. Furthermore, if the \pkg{dbplyr} functionality is employed, \R users can offload the execution of these operations to more powerful RDBMS's. 

\subsection{Extending \pkg{etl}}

The \pkg{etl} package provides tools to speed the development of \pkg{etl}-dependent packages. The \func{create\_etl\_package} function creates a new \R package by calling \func{devtools::create}~\citep{devtools}, while also adding \pkg{etl} to the \cmd{Depends} section of the \cmd{DESCRIPTION} file, and providing the code template shown below, with \val{foo} replaced by \argument{newpkg}. 

\begin{knitrout}
\definecolor{shadecolor}{rgb}{0.969, 0.969, 0.969}\color{fgcolor}\begin{kframe}
\begin{alltt}
\hlstd{proj_dir} \hlkwb{<-} \hlkwd{file.path}\hlstd{(}\hlkwd{tempdir}\hlstd{(),} \hlstr{"newpkg"}\hlstd{)}
\hlkwd{create_etl_package}\hlstd{(proj_dir)}
\end{alltt}

{\ttfamily\noindent\itshape\color{messagecolor}{\#\# Creating package 'newpkg' in '/tmp/Rtmpdcqje1'}}

{\ttfamily\noindent\itshape\color{messagecolor}{\#\# No DESCRIPTION found. Creating with values:}}

{\ttfamily\noindent\itshape\color{messagecolor}{\#\# * Creating `newpkg.Rproj` from template.}}

{\ttfamily\noindent\itshape\color{messagecolor}{\#\# * Adding `.Rproj.user`, `.Rhistory`, `.RData` to ./.gitignore}}

{\ttfamily\noindent\itshape\color{messagecolor}{\#\# * Creating R/etl.R template source file...}}

{\ttfamily\noindent\itshape\color{messagecolor}{\#\# * Adding etl to Depends}}

{\ttfamily\noindent\itshape\color{messagecolor}{\#\# Next:}}

{\ttfamily\noindent\itshape\color{messagecolor}{\#\# Are you sure you want Depends? Imports is almost always the better choice.}}\end{kframe}
\end{knitrout}

A developer can then immediately compile a functioning \R package, which in this case downloads Houston public school district data. The \cmd{default} method for \func{etl\_extract} pulls data provided by the package, which in this case is pointless because there is no such data. Conversely, the \cmd{etl\_foo} method provided in the template illustrates how---in a simple case---a vector of URLs and a call to \func{smart\_download} is sufficient to complete the extract phase. 

\begin{knitrout}
\definecolor{shadecolor}{rgb}{0.969, 0.969, 0.969}\color{fgcolor}\begin{kframe}
\begin{verbatim}
## #' My ETL functions
## #' @import etl
## #' @inheritParams etl::etl_extract
## #' @export
## #' @examples
## #' \dontrun{
## #' if (require(dplyr)) {
## #'   obj <- etl("foo") %>%
## #'     etl_create()
## #' }
## #' }
## 
## etl_extract.etl_foo <- function(obj, ...) {
##   # Specify the URLs that you want to download
##   src <- c("http://www.stat.tamu.edu/~sheather/book/docs/datasets/HoustonChronicle.csv")
## 
##   # Use the smart_download() function for convenience
##   etl::smart_download(obj, src, ...)
## 
##   # Always return obj invisibly to ensure pipeability!
##   invisible(obj)
## }
\end{verbatim}
\end{kframe}
\end{knitrout}

Since the raw data is already in a CSV format, the \cmd{default} methods for \func{etl\_transform} and \func{etl\_load} are sufficient to complete the ETL cycle for this simple example, so there is no need to write \cmd{etl\_foo} methods for these functions. After changing to the root directory of the new package, one can install, load, and use \pkg{newpkg} just like any other. 

\begin{knitrout}
\definecolor{shadecolor}{rgb}{0.969, 0.969, 0.969}\color{fgcolor}\begin{kframe}
\begin{alltt}
\hlstd{devtools}\hlopt{::}\hlkwd{install}\hlstd{()}
\hlkwd{library}\hlstd{(newpkg)}
\hlstd{districts} \hlkwb{<-} \hlkwd{etl}\hlstd{(}\hlstr{"newpkg"}\hlstd{)} \hlopt{%>%}
  \hlkwd{etl_create}\hlstd{()}
\hlstd{districts} \hlopt{%>%}
  \hlkwd{tbl}\hlstd{(}\hlstr{"HoustonChronicle"}\hlstd{)}
\end{alltt}
\end{kframe}
\end{knitrout}

This functionality leverages \pkg{devtools} to allow intermediate \R users with no package development experience (e.g., advanced undergraduate statistics and data science majors) to begin creating \pkg{etl}-dependent \R packages, such as those listed in Section~\ref{sec:PAMDAS}. For more information, please see the ``\href{https://cran.r-project.org/web/packages/etl/vignettes/extending_etl.html}{Extending \pkg{etl}}" vignette~\footnote{\url{https://cran.r-project.org/web/packages/etl/vignettes/extending_etl.html}}.

\subsection{Additional functionality for developers}

The \pkg{etl} package contains several additional functions that are useful for developers. Some of these may eventually be passed upstream to \pkg{DBI}. Briefly,

\begin{itemize}
  \item \func{dbRunScript}: execute a sequence of arbitrary SQL commands. This takes a full SQL script and passes the individual SQL statements to \func{DBI::dbExecute}.
  \item \func{dbWipe}: delete all of the tables in a database
  \item \func{match\_files\_by\_year\_months}, \func{extract\_date\_from\_filename}, and \func{valid\_year\_month} assist with working with dates---specifically in conjunction with files that may encode dates in their names (e.g., \val{201307-citibike-tripdata.zip})
  \item \func{smart\_download} and \func{smart\_upload}: only download and upload files that don't already exist
  \item \func{src\_mysql\_cnf} use the \cmd{\textasciitilde/.my.cnf} configuration file to connect to MySQL
\end{itemize}

\section{Conclusion}
\label{sec:conclusion}

\subsection{Future Work}

The \pkg{etl} package does not solve all problems for those working with medium data. There is considerable room for improving the performance of the \pkg{etl} package itself. First, some of the ETL operations should be parallelizable. In particular, \func{etl\_transform} is a good candidate, since it is always working locally. While in some cases the bottleneck for \func{etl\_extract} will be the speed of the user's Internet connection, in others parallel threads---such as those provided by the \pkg{parallel} package---could significantly improve performance. For \func{etl\_load}, the database engine may not support simultaneous imports to the same table. Second, reading and writing data files to the disk is time-consuming. It is possible that new file formats such as \pkg{feather} could reduce latency~\citep{feather}. Third, the data ends up being stored on disk three times: once in its raw format (hopefully compressed), once as a CSV (uncompressed), and once in the database's native file format (optimized). Importing the compressed files directly into the database may be possible in some cases, but care must be taken to ensure the predictability of these functions. Using symbolic links rather than copying files might also be appropriate in some cases. One can of course use \func{etl\_cleanup} to delete either or both of the first two instances, but perhaps a more streamlined process is possible, at least in some cases. 

The \pkg{etl} package fuels rapid development of dependent packages, even among novice \R developers. We know this because many of the \pkg{etl}-dependent packages referenced above were partially developed by undergraduate students. A broad adoption of these \pkg{etl}-dependent packages and a larger installed user base would increase interest in the project and lead to a more robust infrastructure. We plan to continue this work in the future.

\subsection{Discussion}

As data grow larger and larger, more and more people will need to develop the skills necessary to work with them. Yet there is limited room in the undergraduate curriculum for such training. Moreover, exposing students to truly big data requires expensive technical infrastructure, training, and support that will remain burdensome to many faculty members for the foreseeable future. A more realistic approach towards helping students develop their capacity to work with larger data sets is to focus on medium data (rather than big data). These data are still challenging and will still help students develop their understanding of scalability issues, while at the same time having a much lower barrier to entry for both students and faculty. 

At the same time, producing reproducible research on medium data is more difficult than it is on small data---and many researchers already have a hard time with that. As medium and big data become more prevalent in published research, we must not soften our insistence on reproducibility.

Among educators, interest in exposing statistics students to larger and more complex data is growing. Recent guidelines about undergraduate majors in statistics~\citep{asa-guidelines} and data science~\citep{pcmi2016} endorsed by the American Statistical Association emphasize the necessity of exposing students to such data. \cite{horton2015setting} advocate for discussing medium data as a ``precursor" to big data. However, all of the aforementioned challenges to working with medium data present barriers to statistics educators who are quite comfortable with \R, but may not have sufficient experience with SQL. 

We propose this \pkg{etl} package as a mechanism for facilitating reproducible research on medium data for \R users. This has the dual benefit of lowering barriers to entry (minimal SQL required) for larger and more complex data sets, while simultaneously aiding the reproducibility of any subsequent research. 
Not everyone needs to be a data engineer, but many need to wrangle medium data---\pkg{etl} provides a powerful but simplified interface for the latter.


\bibliographystyle{agsm}
\bibliography{refs}

\newpage

\vspace{3cm}

\begin{center}
{\Large  {\bf Supplementary Materials for: \\ 

\vspace{2cm}

A Grammar for Reproducible and Painless Extract-Transform-Load Operations\\on Medium Data

}}
\end{center}

\newpage

\appendix

\section{Extended discussion of related work}
\label{sec:related}

In this section we summarize the major considerations that make the \pkg{etl} package a progressive step towards reproducible research on medium data for \R users.

\subsection{Reproducible research}

To understand the current challenges we face in conducting reproducible research on PAMDAS, one must start with the notion of literate programming~\citep{knuth1984literate}. In literate programming, source code is woven into an annotated narrative, so that one could read the source code and understand not just the code itself, but also how each piece of code fits into the larger design.  

This idea leads to the notion of \emph{reproducibility} in computational science. \cite{donoho2010invitation} paraphrases~\cite{claerbout1994hypertext}: 
\begin{quotation}
An article about a computational result is advertising, not scholarship. The actual scholarship is the full software environment, code and data, that produced the result.
\end{quotation}

\cite{ioannidis2005most} argues that most published research is false, and while his arguments are \emph{statistical} rather than \emph{computational}, they only help to underscore the importance of computational reproducibility. 

In academia, a diverse set of fields including computer science~\citep{donoho2009reproducible}, economics~\citep{tier2012}, archeology~\citep{marwick2017computational} and neuroscience~\citep{eglen2017toward} are actively debating how they will recognize reproducible research. Organizations like Project TIER (\url{http://www.projecttier.org/}) and the Open Science Framework (\url{https://osf.io/}) provide protocols for conducting reproducible research, while statistics and data science educators are instilling reproducible practices in their students~\citep{baumer2014r}. Top-tier journals like the \textit{Journal of the American Statistical Association} have appointed reproducibility editors~\citep{amstat2016jasa}. 

Thus, while the need for research in all fields to be reproducible is clear, the specifications for what qualifies as reproducible are less clear, and the path towards achieving reproducibility is murkier still. 

\subsection{Medium data}

In the past few years, \emph{big data} has become an omnipresent buzzword that taps into our collective fascination with things that are massive. However, while a few enormous companies (e.g., Google, Facebook, Amazon, Walmart, etc.) generate and analyze truly big data (on the order of \emph{exabytes} (EB), which are equal to 1000 \emph{petabytes} (PB), which are equal to 1000 \emph{terabytes} (TB), which are equal to 1000 \emph{gigabytes} (GB)), most people who analyze data will never interact meaningfully with \href{https://en.wikipedia.org/wiki/Orders_of_magnitude_(data)}{data of that size}. 

Most people will only encounter data that is \emph{small} (a few gigabytes at most). These data fit effortlessly into a computer's memory, and thus the user experiences no challenges related to the data's size. Because a computer can access data in memory at lightning-fast speeds, efficient data analysis algorithms like searching ($O(n)$), sorting ($O(n \log{n})$), and multiplying matrices (e.g., fitting a regression model) ($O(n^{2.376})$~\citep{williams2012multiplying}) will run nearly instantly---even on a laptop.~\footnote{Computer scientists use Big-O notation to describe the running time of algorithms by comparing the order of magnitude of the number of steps the algorithm takes to execute on an input of size $n$. An algorithm that runs in $O(n)$ time is \emph{linear}, in the sense that the amount of time it will take to run is linearly proportional to the size of the input.} Thus, for people working with small data, fundamental computer science concepts like the distinction between hardware and software, algorithmic efficiency, and bus speeds are immaterial.

For the vast majority of us who are unlikely to ever interact meaningfully with truly big data, \emph{medium data} is both a viable solution and an accessible introduction to the challenges of big data~\citep{horton2015setting}. In Table~\ref{tab:data}, we constrast the relative sizes of data from the point of view of a personal computer user. Medium data is on the order of several gigabytes to a few terabytes. These data are large enough that they will not comfortably fit in memory on a personal computer without consequences, making a memory-only application like (vanilla) \R a dubious candidate for data analysis. However, medium data are not so large they won't fit on a single hard disk, making them accessible to a single user without access to a computing cluster. An SQL-based RDBMS remains an appropriate storage and retrieval solution for medium data.

\subsection{Existing challenges}

The fundamental challenge of big data is scalability, but medium data comes with its own challenges. In the end, investment in properly setting up an RDBMS pays off in more efficient analysis.

First, everything with medium data takes a little longer, since the aforementioned algorithms are no longer instantaneous. A single line of code might take one minute to execute instead of a millisecond, but these brief delays compound. Thus, those who employ efficient code and workflows are rewarded for their efforts with shorter execution times. 

Second, a data analyst has to know something about SQL administration in order to set up a database. Many introductory data science courses that teach SQL focus on writing \cmd{SELECT} queries to retrieve data from an existing database---not on writing table schemas and defining keys and indexes~\citep{hardin2015data}. 

Third, getting PAMDAS set up involves often laborious ETL operations. Downloading medium data is not instantaneous and is dependent on the speed of one's Internet connection. Wrangling data is notoriously time-consuming work: reasonable estimates suggest this may occupy as much as 50--80\% of a data scientist's time. 

For these reasons, a responsible data scientist will record their ETL operations in a script. But these scripts are often problematic, ad hoc solutions. Some common problems include:
\begin{description}
  \item[Portability] Shell scripts may not port across operating systems. While Apple's OS X operating system is POSIX-compliant, not all flavors of GNU/Linux are. Microsoft Windows \href{https://en.wikipedia.org/wiki/POSIX#Mostly_POSIX-compliant}{requires additional software} to implement a compatibility layer, and thus any such scripts are not likely to run on Windows without careful modification. 
  \item[Usability] Under time pressure, data scientists are likely to write scripts that work for them, and not necessarily for other people. Their scripts may be idiosyncratic and difficult for another person to use or modify. 
  \item[Version Control] Even if a data scientist uses a formal version control system like \cmd{git} and GitHub, a script that ran when it was written may not run at all points in the future. 
  \item[Languages] ETL scripts may be written in \cmd{bash}, Python, \R, SQL, Perl, PHP, Ruby, Scala, Julia, or any combination of these languages and others. There may be good reasons for mixing different languages but ease of portability decreases with each additional language. 
\end{description}

One recommended solution for bundling ETL scripts for \R users is to create an \R package~\citep{wickham2015r}. Packages provide users with software that extends the core functionality of \R, and often data that illustrates the use of that functionality. \R packages hosted on CRAN---the authoritative central repository---are checked for quality and documentation, helping to ensure their \emph{usability}. Since \R is cross-platform, these packages are \emph{portable}. CRAN itself maintains distinct \emph{versioning}, and while \R packages are mostly written in \R, there are a number of ways in which code from other \emph{languages} can be embedded into an \R package (e.g., \pkg{Rcpp} provides functionality to bundle \cmd{C++} code~\citep{Rcpp}). 

However, by design the types of data that can be contained in an \R package hosted on CRAN are limited. First, packages are designed to be small, so that the amount of data stored in a package is supposed to be less than 5 \emph{megabytes}. Furthermore, these data are static, in that CRAN allows only monthly releases. Alternative package repositories---such as GitHub---are also limited in their ability to store and deliver data that could be changing in real-time to \R users. In Table~\ref{tab:flights} we contrast two different CRAN packages for on-time airline flight data~\citep{nycflights13, hflights}, with an \pkg{etl}-dependent package that allows the user to build their own database of flight data~\citep{airlines}. We note the change in scope that the \pkg{airlines} package allows: whereas the two existing data sets are restricted to small, static data from flights departing two Houston-area airports in 2011, or three New York City-area airports in 2013, respectively, the \pkg{airlines} package covers all domestic flights since 1987 departing from more than 350 airports nationwide, with more data available monthly. 

\begin{table}
  \centering
  \begin{tabular}{cccc}
package       & timespan & airports & size \\
\hline
\pkg{hflights}    &  2011  & IAH, HOU  & 2.1 MB \\ 
\pkg{nycflights13} & 2013 & LGA, JFK, EWR & 4.4 MB \\
\pkg{airlines} & 1987--present & $\approx 350$ &  $> 6$ GB \\
\hline
  \end{tabular}
  \caption{Alternative packaging of on-time flight data from the Bureau of Transportation Statistics in \R. We note that the full scope of flight data is only accessible through the \pkg{airlines} package. \label{tab:flights}}
\end{table}

Many \R packages facilitate the retrieval of data from specific sources. In particular, the rOpenSci group maintains dozens of such packages~\citep{boettiger2015building}. Other popular small CRAN packages that serve as APIs to large data sets include \pkg{tigris}~\citep{tigris} and \pkg{UScensus2010}~\citep{UScensus2010}. While these packages are undoubtedly useful, they are written by many different authors, and the syntax employed across packages varies greatly. In short, there is no consistent ``grammar" (see Section~\ref{sec:grammar}). These packages are peripherals without a core. 

Some dependency approaches do exist. \cite{peng2008statistical} illustrate how a small package for CRAN that interacts with large data repositories not hosted on CRAN could facilitate research in environmental epidemiology. These repositories are maintained by the package author through the use of a second package~\citep{eckel2009interacting}. More recently, the \pkg{drat} package provides a core that facilitates the creation of peripheral packages~\citep{drat}. In this scheme the peripheral packages contain large amounts of data. The major drawback to both of these approaches is the requirement that the researcher maintain the large data repositories. 

\cite{boettiger2015introduction} advocates for the container-based solution Docker as an alternative packaging structure for reproducible research, and more recently Rocker~\citep{boettiger2017introduction}, which provides Docker containers for \R and RStudio. \cite{ccetinkaya2017infrastructure} promote this approach as university instructors. We see \pkg{etl} as fitting nicely into this paradigm, serving to further reduce barriers to reproducibility.

Perhaps the closest competitor to our approach is \pkg{pitchRx}~\citep{pitchRx}, which performs ETL operations for a specific data set---in this case, detailed pitch information from Major League Baseball. Our approach places similar core functionality in the \pkg{etl} package and separates the data-source-specific functionality into small, easy-to-write packages that can be hosted on CRAN. The developer need not maintain any large data repositories---they need only to maintain the small bits of code that interact with the data provider. If, for any reason, the source data changes, \pkg{etl} users still retain copies of the raw data as they downloaded it.  

We imagine that many of these aforementioned packages could be re-factored to have \pkg{etl} as a depedendency.

\section{A toy example}
\label{sec:mtcars}

Here, we illustrate the functionality of the \pkg{etl} package on the built-in \data{mtcars} data set.  

The first step is to instantiate an \cmd{etl} object using the \func{etl} function. We use the \func{etl\_create} function to perform the entire ETL cycle on an object named \obj{my\_cars}. During this process, a local SQLite database is created in a temporary directory, that database is initialized, the \data{mtcars} data is ``downloaded" (i.e., in this case, from memory), transformed, and finally uploaded to that same SQLite database. 

\begin{knitrout}
\definecolor{shadecolor}{rgb}{0.969, 0.969, 0.969}\color{fgcolor}\begin{kframe}
\begin{alltt}
\hlstd{my_cars} \hlkwb{<-} \hlkwd{etl}\hlstd{(}\hlstr{"mtcars"}\hlstd{)} \hlopt{%>%}
  \hlkwd{etl_create}\hlstd{()}
\end{alltt}

{\ttfamily\noindent\itshape\color{messagecolor}{\#\# No database was specified so I created one for you at:}}

{\ttfamily\noindent\itshape\color{messagecolor}{\#\# /tmp/Rtmpdcqje1/file4fd547ea2707.sqlite3}}

{\ttfamily\noindent\itshape\color{messagecolor}{\#\# Initializing DB using SQL script init.sqlite}}

{\ttfamily\noindent\itshape\color{messagecolor}{\#\# Extracting raw data...}}

{\ttfamily\noindent\itshape\color{messagecolor}{\#\# Transforming raw data...}}

{\ttfamily\noindent\itshape\color{messagecolor}{\#\# Loading 6 file(s) into the database...}}\end{kframe}
\end{knitrout}

The object \obj{my\_cars} is both an \cmd{etl\_mtcars} object and a \cmd{src\_dbi} object---and can thus do anything that any other \cmd{src\_dbi} object can do. It also maintains a connection to the SQLite database, has two folders (e.g., \cmd{raw} and \cmd{load}) where it can store files, and knows about a table called \cmd{mtcars} that exists in the SQLite database. 

\begin{knitrout}
\definecolor{shadecolor}{rgb}{0.969, 0.969, 0.969}\color{fgcolor}\begin{kframe}
\begin{alltt}
\hlkwd{class}\hlstd{(my_cars)}
\end{alltt}
\begin{verbatim}
## [1] "etl_mtcars" "etl"        "src_dbi"    "src_sql"    "src"
\end{verbatim}
\begin{alltt}
\hlkwd{summary}\hlstd{(my_cars)}
\end{alltt}
\begin{verbatim}
## files:
##   n     size                 path
## 1 6 0.004 GB  /tmp/Rtmpdcqje1/raw
## 2 6 0.004 GB /tmp/Rtmpdcqje1/load
##       Length Class            Mode       
## con   1      SQLiteConnection S4         
## disco 2      -none-           environment
\end{verbatim}
\begin{alltt}
\hlstd{my_cars}
\end{alltt}
\begin{verbatim}
## dir:  12 files occupying 0.008 GB
## src:  sqlite 3.22.0 [/tmp/Rtmpdcqje1/file4fd547ea2707.sqlite3]
## tbls: atmos, borders, elev, glaciers, mtcars, storms
\end{verbatim}
\end{kframe}
\end{knitrout}

Since \obj{my\_cars} is a \pkg{DBI} data source, the data stored in the SQLite database can be accessed in the usual manner. Here, we compute the average fuel economy for these cars. Note that these computations are performed by SQLite. 

\begin{knitrout}
\definecolor{shadecolor}{rgb}{0.969, 0.969, 0.969}\color{fgcolor}\begin{kframe}
\begin{alltt}
\hlstd{my_cars} \hlopt{%>%}
  \hlkwd{tbl}\hlstd{(}\hlstr{"mtcars"}\hlstd{)} \hlopt{%>%}
  \hlkwd{group_by}\hlstd{(cyl)} \hlopt{%>%}
  \hlkwd{summarize}\hlstd{(}\hlkwc{N} \hlstd{=} \hlkwd{n}\hlstd{(),} \hlkwc{mean_mpg} \hlstd{=} \hlkwd{mean}\hlstd{(mpg))}
\end{alltt}

{\ttfamily\noindent\color{warningcolor}{\#\# Warning: Missing values are always removed in SQL.\\\#\# Use `AVG(x, na.rm = TRUE)` to silence this warning}}\begin{verbatim}
## # Source:   lazy query [?? x 3]
## # Database: sqlite 3.22.0 [/tmp/Rtmpdcqje1/file4fd547ea2707.sqlite3]
##     cyl     N mean_mpg
##   <int> <int>    <dbl>
## 1     4    11     26.7
## 2     6     7     19.7
## 3     8    14     15.1
\end{verbatim}
\end{kframe}
\end{knitrout}

The \obj{my\_cars} object itself occupies very little of \R's memory. 

\begin{knitrout}
\definecolor{shadecolor}{rgb}{0.969, 0.969, 0.969}\color{fgcolor}\begin{kframe}
\begin{alltt}
\hlstd{my_cars} \hlopt{%>%}
  \hlkwd{object.size}\hlstd{()} \hlopt{%>%}
  \hlkwd{print}\hlstd{(}\hlkwc{units} \hlstd{=} \hlstr{"Kb"}\hlstd{)}
\end{alltt}
\begin{verbatim}
## 3.2 Kb
\end{verbatim}
\end{kframe}
\end{knitrout}

\section{Benchmarking}
\label{sec:benchmark}

Recall that in Section~\ref{sec:bikes} we created a \cmd{tbl\_dbi} called \obj{trips} that is connected to a database table of Citi Bike trip rentals. In this example we illustrate how the ability of \pkg{dplyr} to offload certain computations to SQL can result in marked performance improvements, even on the same computer. 

\begin{knitrout}
\definecolor{shadecolor}{rgb}{0.969, 0.969, 0.969}\color{fgcolor}\begin{kframe}
\begin{alltt}
\hlkwd{class}\hlstd{(trips)}
\end{alltt}
\begin{verbatim}
## [1] "tbl_dbi"  "tbl_sql"  "tbl_lazy" "tbl"
\end{verbatim}
\end{kframe}
\end{knitrout}

Previously, we used the following pipeline to compute the number of unique combinations of stations, days, and hours in the month of September 2013. In the code below, we make use of the lazy evaluation design of \pkg{dplyr} to push the computation to MySQL. Note that the functions in uppercase are MySQL functions---not \R functions. The \func{collect} verb is applied only after the database is queried so that \R can count the number of resulting rows. Because MySQL is good at doing this type of operation, and only $167,258$ rows of data are sent from MySQL to \R, this computation takes only a few seconds. 

\begin{knitrout}
\definecolor{shadecolor}{rgb}{0.969, 0.969, 0.969}\color{fgcolor}\begin{kframe}
\begin{alltt}
\hlkwd{system.time}\hlstd{(}
\hlstd{trips_sept} \hlkwb{<-} \hlstd{trips} \hlopt{%>%}
  \hlkwd{filter}\hlstd{(}\hlkwd{YEAR}\hlstd{(start_time)} \hlopt{==} \hlnum{2013}\hlstd{)} \hlopt{%>%}
  \hlkwd{group_by}\hlstd{(start_station_id,} \hlkwd{DAY}\hlstd{(start_time),} \hlkwd{HOUR}\hlstd{(start_time))} \hlopt{%>%}
  \hlkwd{summarize}\hlstd{(}\hlkwc{N} \hlstd{=} \hlkwd{n}\hlstd{(),}
            \hlkwc{num_stations} \hlstd{=} \hlkwd{COUNT}\hlstd{(}\hlkwd{DISTINCT}\hlstd{(start_station_id)),}
            \hlkwc{num_days} \hlstd{=} \hlkwd{COUNT}\hlstd{(}\hlkwd{DISTINCT}\hlstd{(}\hlkwd{DAYOFYEAR}\hlstd{(start_time))))} \hlopt{%>%}
  \hlkwd{collect}\hlstd{()}
\hlstd{)}
\end{alltt}
\begin{verbatim}
##    user  system elapsed 
##   0.309   0.000   1.750
\end{verbatim}
\begin{alltt}
\hlkwd{nrow}\hlstd{(trips_sept)}
\end{alltt}
\begin{verbatim}
## [1] 167258
\end{verbatim}
\end{kframe}
\end{knitrout}

Conversely, we can use the \pkg{lubridate} package for assistance with dates, and the \func{collect} function to bring the data into \R for summarization. Note here that only the \func{filter} operation is actually performed by MySQL, while the rest of the operations are performed in \R. 

\begin{knitrout}
\definecolor{shadecolor}{rgb}{0.969, 0.969, 0.969}\color{fgcolor}\begin{kframe}
\begin{alltt}
\hlkwd{library}\hlstd{(lubridate)}
\hlkwd{system.time}\hlstd{(}
\hlstd{trips_sept} \hlkwb{<-} \hlstd{trips} \hlopt{%>%}
  \hlkwd{filter}\hlstd{(}\hlkwd{YEAR}\hlstd{(start_time)} \hlopt{==} \hlnum{2013}\hlstd{)} \hlopt{%>%}
  \hlkwd{collect}\hlstd{()} \hlopt{%>%}
  \hlkwd{group_by}\hlstd{(start_station_id,} \hlkwd{day}\hlstd{(start_time),} \hlkwd{hour}\hlstd{(start_time))} \hlopt{%>%}
  \hlkwd{summarize}\hlstd{(}\hlkwc{N} \hlstd{=} \hlkwd{n}\hlstd{(),}
            \hlkwc{num_stations} \hlstd{=} \hlkwd{n_distinct}\hlstd{(start_station_id),}
            \hlkwc{num_days} \hlstd{=} \hlkwd{n_distinct}\hlstd{(}\hlkwd{yday}\hlstd{(start_time)))}
\hlstd{)}
\end{alltt}
\begin{verbatim}
##    user  system elapsed 
##  28.461   0.855  29.322
\end{verbatim}
\begin{alltt}
\hlkwd{nrow}\hlstd{(trips_sept)}
\end{alltt}
\begin{verbatim}
## [1] 167258
\end{verbatim}
\end{kframe}
\end{knitrout}

This latter method is much slower since it has to transfer more than 1 million rows of data from MySQL to \R, instead of only $167,258$. The delay with the second method is noticeable enough to start a conversation with students about scalability. 

\section{Using Amazon RDS}
\label{sec:amazon}

In this section we provide a brief tutorial explaining how to set up a medium database of taxi trip information on Amazon RDS (a cloud-based service) and populate it. 

\begin{figure}
  \centering
  \includegraphics[width=0.8\textwidth]{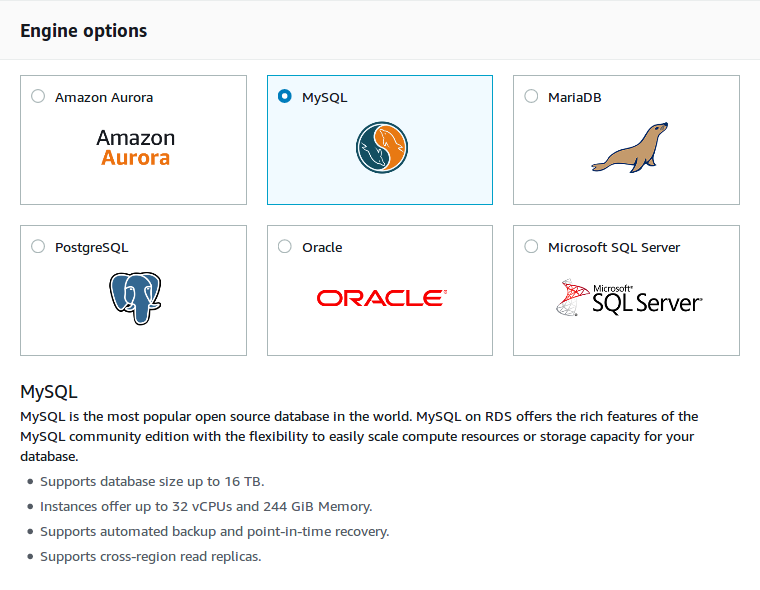}
  \caption{Amazon RDS}
  \label{fig:amazon_engine}
\end{figure}

First, you must set up an Amazon Web Services account at \url{https://aws.amazon.com/rds/}. Our goal is to launch a new relational database service instance. In this example we will create a MySQL database that uses the Free Usage Tier (to avoid fees). In Figure~\ref{fig:amazon_engine}, we show how to select the MySQL engine from among the available options. 

\begin{figure}
  \centering
  \includegraphics[width=0.8\textwidth]{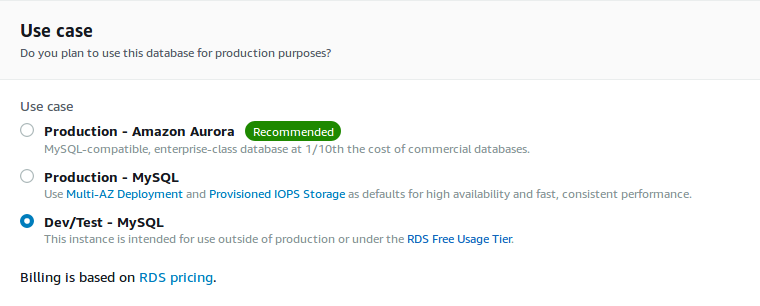}
  \caption{Amazon RDS}
  \label{fig:amazon_use}
\end{figure}

Since we are simply testing this service, we select the ``Dev/Test" usage case, which is the only one that is available under the Free Usage Tier (see Figure~\ref{fig:amazon_use}).

\begin{figure}
  \centering
  \includegraphics[width=0.8\textwidth]{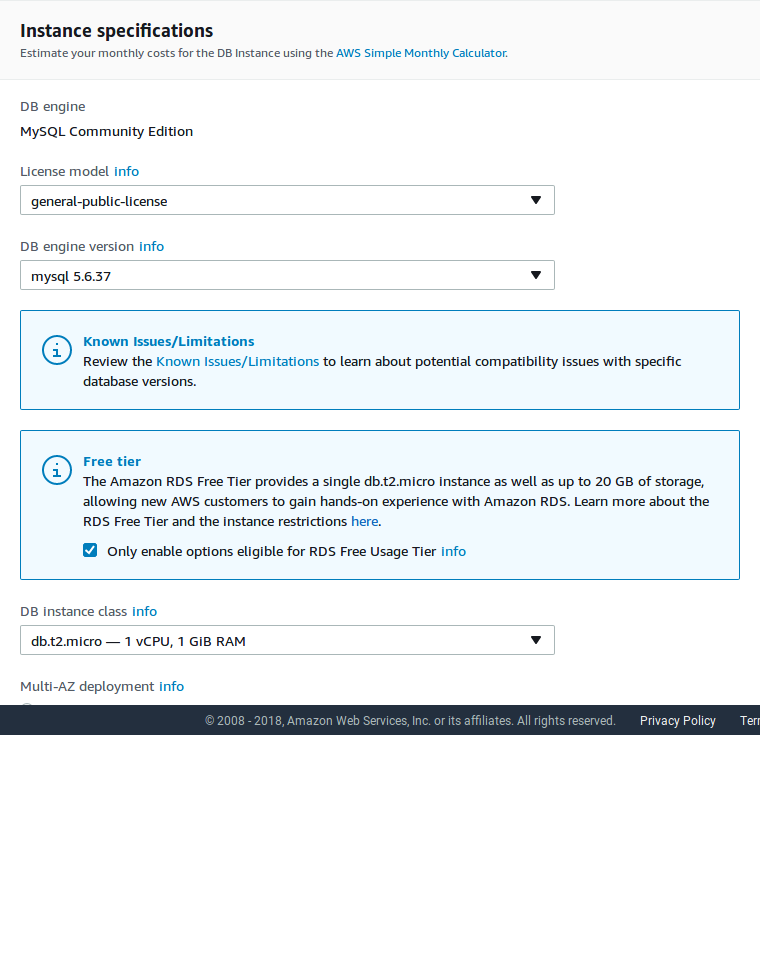}
  \caption{Amazon RDS}
  \label{fig:amazon_db}
\end{figure}

Next, in Figure~\ref{fig:amazon_db} we allocate only minimal resources to this database instance. The \cmd{db.t2.micro} instance has only 1 CPU and 1 gigabyte of memory. This is the only allowable configuration in the Free Usage Tier. 

\begin{figure}
  \centering
  \includegraphics[width=0.8\textwidth]{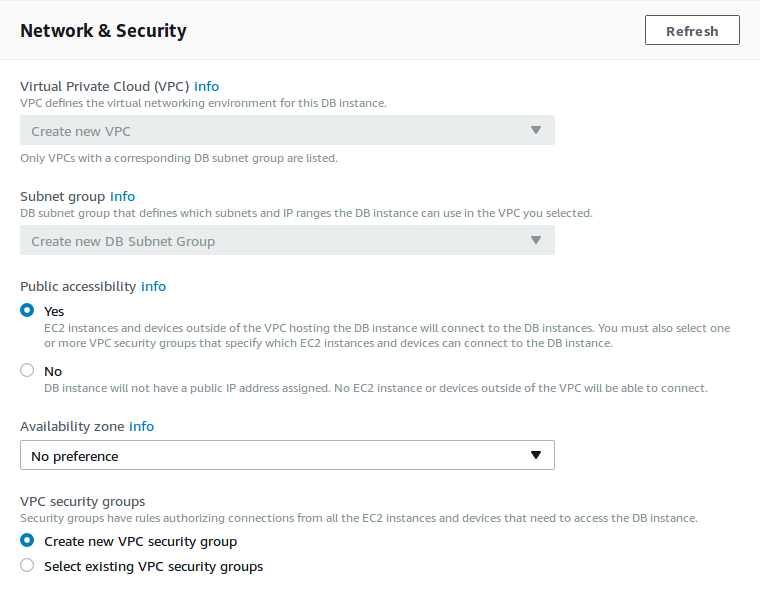}
  \caption{Amazon RDS}
  \label{fig:amazon_security}
\end{figure}

In Figure~\ref{fig:amazon_security}, we elect to make our database publicly accessible. This is an important deviation from the default, which is to restrict access to a Virtual Private Cloud. Without selecting ``Yes" here, we would not be able to connect to our database from our \R client. Please consult the documentation on Amazon in order to fully understand your security settings. Note also that by default, public access is only granted from \emph{your} IP address. 

In the next step, we set up a username, password, and schema. These are specific to the MySQL instance on our cloud-based database server. 
After accepting all of the default options on the remaining screens, our instance will launch. This process creates a virtual MySQL server that is running on Amazon's servers. The hostname for that server is shown in your Instance dashboard under ``Endpoint". 

\begin{knitrout}
\definecolor{shadecolor}{rgb}{0.969, 0.969, 0.969}\color{fgcolor}\begin{kframe}
\begin{alltt}
\hlstd{host} \hlkwb{<-} \hlstr{"etl-test.cdc7tgkkqd0n.us-east-1.rds.amazonaws.com"}
\end{alltt}
\end{kframe}
\end{knitrout}

If we didn't set up a schema on the MySQL server called \charvec{nyctaxi} already, we can create one using the Terminal tab available in RStudio. Be sure to use the credentials for the MySQL instance that you specified. 

\begin{knitrout}
\definecolor{shadecolor}{rgb}{0.969, 0.969, 0.969}\color{fgcolor}\begin{kframe}
\noindent
\ttfamily
\hlstd{mysql\ }\hlopt{{-}}\hlstd{h\ etl{-}test.cdc7tgkkqd0n.us{-}east{-}1.rds.amazonaws.com\ }\hlopt{{-}}\hlstd{u\ bbaumer\ }\hlopt{{-}}\hlstd{p\ }\hlopt{{-}}\hlstd{e\ }\hlstr{"CREATE\ DATABASE\ IF\ NOT\ EXISTS\ nyctaxi;"}\hlstd{}\hspace*{\fill}
\mbox{}
\normalfont
\end{kframe}
\end{knitrout}

Finally, we load the \pkg{nyctaxi} package and connect to our database instance. 

\begin{knitrout}
\definecolor{shadecolor}{rgb}{0.969, 0.969, 0.969}\color{fgcolor}\begin{kframe}
\begin{alltt}
\hlkwd{library}\hlstd{(nyctaxi)}
\hlstd{db_rds} \hlkwb{<-} \hlkwd{src_mysql}\hlstd{(}\hlkwc{dbname} \hlstd{=} \hlstr{"nyctaxi"}\hlstd{,}
                    \hlkwc{host} \hlstd{=} \hlstr{"etl-test.cdc7tgkkqd0n.us-east-1.rds.amazonaws.com"}\hlstd{,}
                    \hlkwc{user} \hlstd{=} \hlstr{"bbaumer"}\hlstd{,}
                    \hlkwc{password} \hlstd{=} \hlstr{"xxxxxxxx"}\hlstd{)}
\end{alltt}
\end{kframe}
\end{knitrout}

The \pkg{etl} grammar now allows us to easily populate the database. 

\begin{knitrout}
\definecolor{shadecolor}{rgb}{0.969, 0.969, 0.969}\color{fgcolor}\begin{kframe}
\begin{alltt}
\hlstd{rides} \hlkwb{<-} \hlkwd{etl}\hlstd{(}\hlstr{"nyctaxi"}\hlstd{,} \hlkwc{db} \hlstd{= db_rds,} \hlkwc{dir} \hlstd{=} \hlstr{"~/dumps/nyctaxi"}\hlstd{)}
\hlstd{rides} \hlopt{%>%}
  \hlkwd{etl_update}\hlstd{(}\hlkwc{years} \hlstd{=} \hlnum{2014}\hlstd{,} \hlkwc{months} \hlstd{=} \hlnum{3}\hlstd{)}
\end{alltt}
\end{kframe}
\end{knitrout}

\end{document}